\documentclass[twocolumn,aps,floats,superscriptaddress]{revtex4}

\usepackage{graphicx,epsfig}
\usepackage{times}
\usepackage{graphics,dcolumn,bm,float}
\usepackage{amssymb,amsmath,rotate,color}
\usepackage[normalem]{ulem}
\usepackage{soul}
\usepackage{amsmath}
\usepackage[colorlinks,citecolor = blue, linkcolor = blue, urlcolor  = blue, anchorcolor = blue, breaklinks = true]{hyperref}

\begin{document}
\title{Electron scattering in a superlattice of line defects on the surface of topological insulators}
\author{H. Dehnavi}
\affiliation{Department of Physics, North Tehran Branch, Islamic Azad University, Tehran, Iran}
\author{A. A. Masoudi}
\altaffiliation{masoudi@alzahra.ac.ir}\affiliation{Department of Physics, Alzahra University, Tehran 1993891167, Iran}
\affiliation{Department of Physics, North Tehran Branch, Islamic Azad University, Tehran, Iran}
\author{M. Saadat}
\affiliation{Department of Physics, Shahid Rajaee Teacher Training University, Tehran, Iran}
\author{H. Ghadiri}
\affiliation{Department of Physics, North Tehran Branch, Islamic Azad University, Tehran, Iran}
\author{A. Saffarzadeh}
\altaffiliation{Author to whom correspondence should be addressed. Electronic mail: asaffarz@sfu.ca}
\affiliation{Department of Physics, Payame Noor University, P.O. Box 19395-3697 Tehran, Iran}
\affiliation{Department of Physics, Simon Fraser University, Burnaby, British Columbia, Canada V5A 1S6}
\date{\today}

\begin{abstract}
The electron scattering from periodic line defects on the surface of topological insulators with hexagonal warping effect is investigated theoretically by means of a transfer matrix method. The influence of surface line defects, acting as structural ripples on propagation of electrons are studied in two perpendicular directions due to the asymmetry of warped energy contour under momentum exchange. The transmission profiles and the details of resonant peaks which vary with the number of defects and the strength of their potentials are strongly dependent on the direction in which the line defects extend. At low energies, the quantum interference between the incident and reflected propagating electrons has the dominant contribution in transmission resonances, while at high energies the multiple scattering processes on the constant-energy contour also appear because of the strong warping effect. By increasing the spatial separation between the line defects, the minimum value of electrical conductance remains significantly high at low incident energies, while the minimum value may approach zero at high energies as the number of defects is increased. Our findings suggest that the potential ripples on the surface of topological insulators can be utilized to control the local electronic  properties of these materials.

\end{abstract}
\maketitle 

\section{INTRODUCTION}

Three-dimensional (3D) topological insulators have attracted a great deal of attention in the fields of condensed matter physics and materials science due to their scientific importance as a novel quantum state and their spintronic applications \cite{Hasan2010,Qi2010,Qi2011}. The peculiar behavior of these materials in general, with an insulating bulk energy gap and gapless surface states is originated from the strong spin-orbit coupling and time-reversal symmetry \cite{Kane-PRL1,Kane-PRL2,Fu-PRL2007,Fu-PRB2007}. In other words, the conductive surface states in such materials are preserved, as long as the time reversal symmetry remains unbroken \cite{Moore2007-PRB,nagaosa2007-SCI,moore2009-NAT}. 

The scattering of electron wave function by impurities on the surface of topological insulators plays a dominant role in electronic properties of these materials \cite{Xu-NaCo2017,Miao2018,Liu-PRL2009,Zyuzin-PRB2007}. The existence of impurities in the presence of electric field and spin-orbit interaction, may cause a spin-dependent dispersion of electrons. In fact, the non-polarized spin beams deviate to the edges of the surface, and hence, the scattering amplitude becomes spin-dependent. This effect was empirically observed in thin layers of GaAs and InGaAs \cite{kato2004-SCI}. It has been reported that the impurities can also reshape the surface Dirac cone of the 3D topological insulator Bi$_{2}$Se$_{3}$ \cite{Miao2018}. On the other hand, magnetic impurities can open up a local energy gap, suppress the local density of states, and induce an RKKY interaction in the system \cite{Liu-PRL2009,Zyuzin-PRB2007}.

Besides impurities, surface defects and ripples in these materials can also scatter Dirac electrons. Bismuth-based topological insulators, such as Bi$_2$Se$_3$ and Bi$_2$Te$_3$ \cite{zhang2009-NP} are inexpensive materials that can be prepared easily with excellent conductivity on their surfaces \cite{xia2009-Nat}. Okada et al. \cite{okada2012-NAT}, created a series of ripples on the surface of Bi$_2$Te$_3$, suggesting that two-dimensional (2D) ripples are an efficient method for controlling the properties of Dirac fermions in topological insulators. Such local ripples are imposed by periodic buckling which can be created during the sample growth or by inducing strain through a piezoelectric crystal \cite{okada2012-NAT}.

The one-dimensional periodic potentials (superlattice) on the surface of materials are responsible for trapping the surface electrons and shifting the energy bands.  In this regard, regular line defects (local ripples) on the surface of topological insulators, can be modeled by delta-function potentials \cite{an2012-PRB,zhang2012-PRB,li2013-JAP}. Using a quantum mechanical approach, An \textit{et al.}, \cite{an2012-PRB} investigated the surface states scattering from a potential barrier (single line defect) in Bi$_{2}$Te$_{3}$ with hexagonal warping effect. They showed that the transmission of incident electrons can be perfect when the line defect is along $\Gamma$M, while there is a finite probability for electrons to be reflected when the defect runs along $\Gamma$K. As the Fermi energy is increased, the warping parameter becomes responsible for variation of constant-energy contour (CEC) from a circular shape to a hexagon and then to a snowflake with sharp tips along the $\Gamma$M directions. The warped CEC is modulated by external magnetic field \cite{Chiba2017} or magnetic proximity effect \cite{Arabikhah2019}. Moreover, the warping strength may considerably enhance the electric conductance at high energies relative to the Dirac point as a result of the induced-transport channels by the snowflake shape of the CEC \cite{Arabikhah2019}. The influence of two and three delta-function potentials along $\Gamma$K on the transmission of incident electrons in Bi$_{2}$Te$_{3}$ has also been reported \cite{li2013-JAP}. The results showed that the electric conductance oscillates with delta-function potential strength and that the electronic transport on the surface of the topological insulator can be controlled by structural parameters.

The aim of this paper is to explore the influence of structural ripples on electronic transport in topological insulators with warped surface states. We consider a series of local potential barriers in both $\Gamma$K and $\Gamma$M directions. We obtain transmission and reflection coefficients and also conductance of incident electrons through the superlattice of potential for different structural parameters, namely the number of potential barriers, potential strength, the distance between barriers, incident electron energy, and the angle of incidence.

\section{METHOD}
We consider the 3D topological insulator Bi$_{2}$Te$_{3}$ with strong warping effects and a Dirac cone on its surface. By tuning the Fermi level on the surface states, one can ignore the interaction between surface and bulk states \cite{chen2009-SCI}. The single-particle Hamiltonian for the surface electrons with hexagonal warped states can be expressed as 
  \begin{eqnarray}\label{Hamiltonian}
 H(p_x,p_y)&=&v(p_x\sigma_y-p_y\sigma_x)+\lambda(p_x^3-3p_xp_y^2)\sigma_z, 
\end{eqnarray}                  
where $p_x=-i\hbar\partial_x$ and $p_y=-i\hbar\partial_y$ are the 2D momentum operators of surface electrons, $\sigma_j(j=x,y,z)$ are the Pauli spin matrices, $v=2.55$ eV$\cdot$\AA\ is the Fermi velocity, and $\lambda=250$ eV$\cdot$\AA$^3$ is the warping parameter \cite{chen2009-SCI,fu2009-PRL}. The energy eigenvalues and eigenstates of the Hamiltonian in units of $\hbar=1$ are given as \cite{lee2009-PRB,basak2011-PRB} 
\begin{eqnarray}\label{EK}
E_{\bf k}&=&\pm\sqrt{\lambda^{2}(k_{x}^{3}-3k_{x}k_{y}^{2})^{2}+v^{2}(k_{x}^{2}+k_{y}^{2})},\label{eigenvalue}\\
\psi_{\bf k}({\bf r})&=&\dfrac{1}{2\pi N_{\bf k}} \begin{pmatrix} \phi_{\bf k}+E_{\bf k}\vspace{3 mm}\\ v(ik_{x}-k_{y}) \end{pmatrix}
\mathrm{e}^{i(k_{x}x+k_{y}y)},
\end{eqnarray}
where $N_{\bf k}$ is a normalization factor and $\phi_{\bf k}=\lambda k_{x}(k_{x}^{2}-3k_{y}^{2})$ \cite{an2012-PRB}. 
In the case of electron propagation in $\Gamma$K($x$) direction, we introduce $\widetilde{\psi}_{\bf k}({\bf r})$ as 
\begin{equation}\label{psitilde}
\widetilde{\psi}_{\bf k}({\bf r})=\dfrac{1}{\sqrt{\mid v_x({\bf k})\mid}}\psi_{\bf k}({\bf r}).
\end{equation} 

The expectation value of current operator in that direction, $\hat{v}_x$, is obtained as $\langle \widetilde{\psi}_{\bf k}|\hat{v}_x|\widetilde{\psi}_{\bf k}\rangle=\frac{v_x({\bf k})}{|v_x({\bf k})|}$ where $v_x({\bf k})=({\partial E_{\bf k}}/{\partial k_{x}})_{k_{y}}$. This leads to a more physical expression for the current conservation \cite{an2012-PRB}. Note that the dispersion relation $E_{\bf k}$ is dependent on warping parameter which manifests itself as interesting features, shown in Fig. 1. We can see that the CEC deforms from circular to hexagon and then to snowflake shapes, as the Fermi energy is increased. Therefore, we expect different transport spectra in $k_{x}$ and $k_{y}$ directions that will be discussed below \cite{chen2009-SCI,fu2009-PRL,kuroda2010-PRL}.

For the fixed energy $E_{\bf k}$ and $k_y$ values and in the case of $E_{\bf k}>E_{cx}\simeq 373$ meV, in a special range of $k_{y}$ where CEC has inflection points (see Fig. 1), Eq. (2) has six real roots for $k_{x}$ which are symmetric (out of this range we have two real and four complex roots). The solution indicates that in the presence of a delta-function potential barrier, an incoming electron wave from the left will have three propagating reflected waves towards the left, and three propagating transmitted waves towards the right which are the manifestation of warping effect. 
We call them $k_{x_\alpha}$ ($\alpha=1,2,\cdots,6$) with $k_{x_1}>0$, $k_{x_2}=-k_{x_1}$, $k_{x_5}>0$, $k_{x_6}=-k_{x_5}$ corresponding to electron-like propagation because for these roots $k_{x_\alpha}v_x(k_{x_\alpha},k_y)>0$. Also, $k_{x_3}<0$ and $k_{x_4}=-k_{x_3}$ correspond to hole-like propagation because for these roots $k_{x_\alpha}v_x(k_{x_\alpha},k_y)<0$. If $E_{\bf k}<E_{cx}$ we will have two real roots $k_{x_1}>0$, $k_{x_2}=-k_{x_1}$ and four complex roots $k_{x_3}$, $k_{x_4}=-k_{x_3}$, $k_{x_5}$ and $k_{x_6}=-k_{x_5}$ so that $\text{Im}(k_{x_3})>0$, $\text{Im}(k_{x_4})<0$, $\text{Im}(k_{x_5})>0$ and $\text{Im}(k_{x_6})<0$. There is a similar discussion when $E_{\bf k}$ and $k_x$ values are known. In this case, the characteristics equation in terms of $k_y$ is of order four. For $E_{\bf k}>E_{cy}\simeq 180$ meV, in a special range of $k_{x}$, four $k_y$ values are real; two propagating modes are electron like and two modes are hole like, indicating that an incoming electron wave from the left will have two propagating reflected waves towards the left, and two propagating transmitted waves towards the right in the presence of a delta-function potential barrier. For $E_{\bf k}< E_{cy}$, two modes decay in far distances and the other two modes are propagating \cite{an2012-PRB,li2013-JAP}.

We distinguish different values of $k_x$ and $k_y$ by index ($k_{\alpha x}, k_{\alpha y}$) and for future purpose we write Eq. (\ref{psitilde}) in the following form
\begin{equation}
\widetilde\psi_{k_\alpha}({\bf r})=\begin{pmatrix} c_{k_\alpha}^{(1)}\vspace{2 mm}\\ c_{k_\alpha}^{(2)} \end{pmatrix}\mathrm{e}^{i(k_{\alpha x}x+k_{\alpha y}y)},
\end{equation}
\begin{figure}
\center\includegraphics[width=1.05\linewidth]{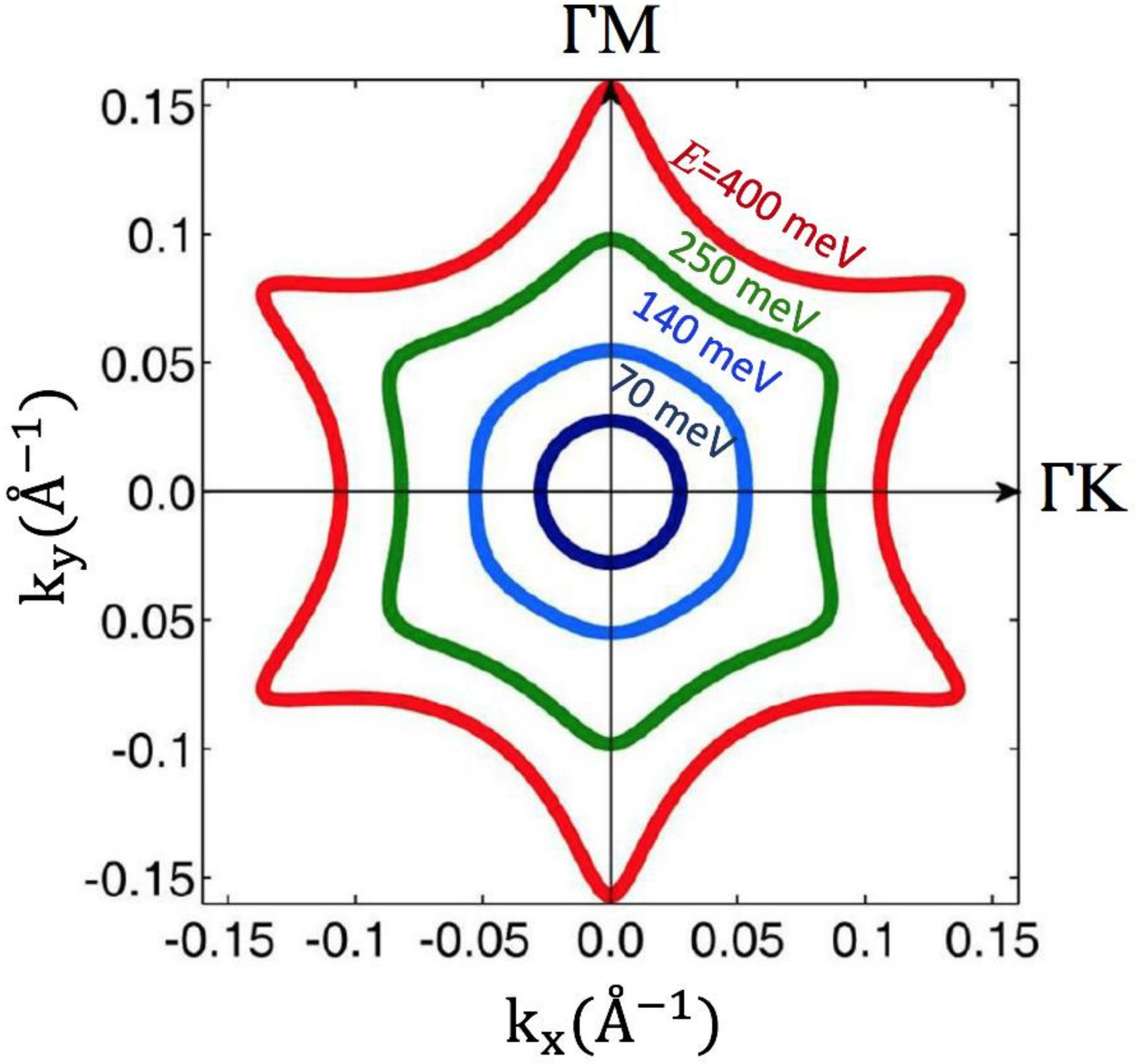}
\caption{The constant-energy contours of 2D Bi$_{2}$Te$_{3}$ at different energies measured from the Dirac point.} 
\label{figure1}
\end{figure}
\begin{figure}
\center\includegraphics[width=0.9\linewidth]{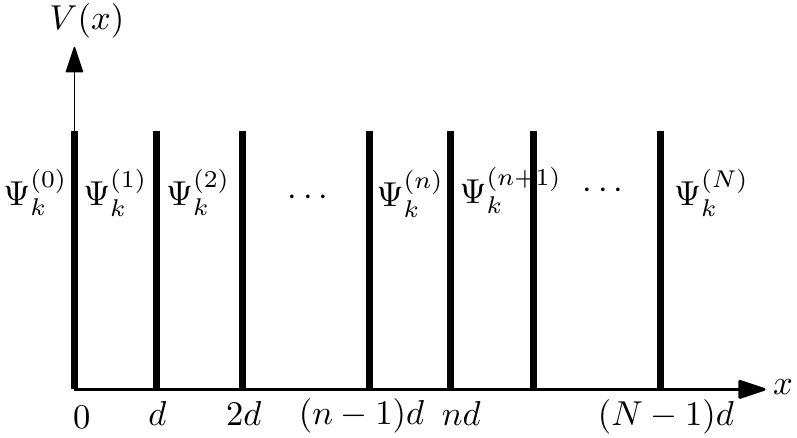}
\caption{Schematics of the superlattice potential consisting of $N$ delta-function potential barriers (line defects) along $x$ direction which represent structural ripples on the surface of topological insulators Bi$_{2}$Te$_{3}$.}
\label{figure2}
\end{figure}

\subsection{A superlattice of line defects along $\Gamma$K($x$)}

First, we model structural ripples as a regular array of $N$ delta-function potentials (line defects, as shown in Fig. 2) on the surface of topological insulator Bi$_{2}$Te$_{3}$ along $x$ direction ($\Gamma$K direction in {\bf k}-space, shown in Fig. 1), namely the ripples are parallel to $y$ axes. 
Note that the advantage of delta-function potentials over other forms of scattering potentials, such as rectangular barriers, is that electrons on upper and lower surface potentials on both sides of the defects have the same surface potential and there is no potential difference across each line defect. Therefore, the model Hamiltonian (1) remains valid, because electron energy in all scattering regions does not violate the well-defined low-energy region \cite{an2012-PRB}. The potential function of these defects is expressed as 
\begin{equation}\label{Vx}
V(x)=\sum_{n=0}^{N-1}U\delta(x-nd).
\end{equation}
Here, $U$ is the strength of delta-function potential and $d$ (lattice constant) is the distance between two adjacent potentials. 
It is worth mentioning that an experimental observation \cite{okada2012-NAT} on the surface of Bi$_{2}$Te$_{3}$ has shown that a periodic modulation of atomic positions (buckling) perpendicular to the surface is an efficient method for manipulating local electronic properties. In this regard, a physical buckling was created by applying the periodic bias voltage 300 mV on the surface. Such a bias voltage can be modeled as a potential barrier with height 0.3 eV and width 1 \AA, that is $U=0.3$ eV$\cdot$\AA. Note that in the present model, the parameter U can take any values as long as electron energy remains in the well-defined low-energy region \cite{an2012-PRB}. 

The electron wave function in the region of $(n-1)d<x<nd$ is given by
\begin{equation}\label{Psi}
\Psi_{\bf k}^{(n)}({\bf r})=\sum_{\alpha=1}^6 a_{\alpha}^{(n)}\widetilde{\psi}_{k_{\alpha}}({\bf r}),
\end{equation}
that is the solution of eigenvalue equation $\{H(-i\partial_{x},k_{y})+\sum_{n=0}^{N-1}U\delta(x-nd)\}\Psi_{\bf k}^{(n)}({\bf r})=E\Psi_{\bf k}^{(n)}({\bf r})$.
Therefore, the boundary condition at $x=nd$ would be in the following form
\begin{eqnarray}\label{boundry1}
&&\Psi_{\bf k}^{(n)}({\bf r}){|_{(x=nd,y)}}=\Psi_{\bf k}^{(n+1)}({\bf r}){|_{(x=nd,y)}}, \nonumber\\
&&\partial_{x}\Psi_{\bf k}^{(n)}({\bf r}){|_{(x=nd,y)}}=\partial_{x}\Psi_{\bf k}^{(n+1)}({\bf r}){|_{(x=nd,y)}},\\
&&\partial_{x}^2(\Psi_{\bf k}^{(n+1)}({\bf r})-\Psi_{\bf k}^{(n)}({\bf r})){|_{(x=nd,y)}}=i\eta\sigma_{z}\Psi_{\bf k}^{(n)}({\bf r}){|_{(x=nd,y)}},\nonumber
\end{eqnarray}
where $\eta=\frac{U}{v}$ \cite{an2012-PRB}. Note that the above eigenvalue equation is a third-order partial differential equation with respect to $x$. The first and second conditions above are the continuities of the wave function and its first derivative, respectively, at $x=nd$. The third condition results from the discontinuity of the second derivative of the wave function due to the delta-function potential and can be obtained by integrating the above eigenvalue equation between $x=nd+0^{-}$ and $x=nd+0^{+}$. After imposing the boundary conditions, we obtain the following relation between wave-function coefficients $a_{\alpha}^{(n)}$ and $a_{\alpha}^{(n+1)}$
\begin{equation}\label{transfer}
a_{\alpha}^{(n+1)}=\sum_{\beta=1}^6 T^{(n)}_{\alpha\beta}a_{\beta}^{(n)}, 
\end{equation}
where $T^{(n)}=(Z_n^{-1}A^{-1}BZ_n)$ is the transfer matrix from $n^{\text{th}}$ defect to $(n+1)^{\text{th}}$ defect. Here, $A$, $B$ and $Z_n$ are $6\times 6$ matrices which are given as
\begin{widetext}
\begin{eqnarray}
A=
\begin{pmatrix}
c_{k_1}^{(1)}&c_{k_3}^{(1)}&c_{k_5}^{(1)}&c_{k_2}^{(1)}&c_{k_4}^{(1)}&c_{k_6}^{(1)}\vspace{3 mm}\\c_{k_1}^{(2)}&c_{k_3}^{(2)}&c_{k_5}^{(2)}&c_{k_2}^{(2)}&c_{k_4}^{(2)}&c_{k_6}^{(2)}\vspace{3 mm}
\\(ik_{x_{1}})c_{k_1}^{(1)}&(ik_{x_{3}})c_{k_3}^{(1)}&(ik_{x_{5}})c_{k_5}^{(1)}&(ik_{x_{2}})c_{k_2}^{(1)}&(ik_{x_{4}})c_{k_4}^{(1)}&(ik_{x_{6}})c_{k_6}^{(1)}\vspace{3 mm}
\\(ik_{x_{1}})c_{k_1}^{(2)}&(ik_{x_{3}})c_{k_3}^{(2)}&(ik_{x_{5}})c_{k_5}^{(2)}&(ik_{x_{2}})c_{k_2}^{(2)}&(ik_{x_{4}})c_{k_4}^{(2)}&(ik_{x_{6}})c_{k_6}^{(2)}\vspace{3 mm}
\\(k_{x_{1}}^{2}+iv\eta/\lambda)c_{k_1}^{(1)}&(k_{x_{3}}^{2}+iv\eta/\lambda)c_{k_3}^{(1)}&(k_{x_{5}}^{2}+iv\eta/\lambda)c_{k_5}^{(1)}&(k_{x_{2}}^{2}+iv\eta/\lambda)c_{k_2}^{(1)}&(k_{x_{4}}^{2}+iv\eta/\lambda)c_{k_4}^{(1)}&(k_{x_{6}}^{2}+iv\eta/\lambda)c_{k_6}^{(1)}\vspace{3 mm}
\\(k_{x_{1}}^{2}-iv\eta/\lambda)c_{k_1}^{(2)}&(k_{x_{3}}^{2}-iv\eta/\lambda)c_{k_3}^{(2)}&(k_{x_{5}}^{2}-iv\eta/\lambda)c_{k_5}^{(2)}&(k_{x_{2}}^{2}-iv\eta/\lambda)c_{k_2}^{(2)}&(k_{x_{4}}^{2}-iv\eta/\lambda)c_{k_4}^{(2)}&(k_{x_{6}}^{2}-iv\eta/\lambda)c_{k_6}^{(2)}
\end{pmatrix},
\end{eqnarray}

\begin{equation}
B=
\begin{pmatrix}
c_{k_1}^{(1)}&c_{k_3}^{(1)}&c_{k_5}^{(1)}&c_{k_2}^{(1)}&c_{k_4}^{(1)}&c_{k_6}^{(1)}\vspace{3 mm}\\c_{k_1}^{(2)}&c_{k_3}^{(2)}&c_{k_5}^{(2)}&c_{k_2}^{(2)}&c_{k_4}^{(2)}&c_{k_6}^{(2)}\vspace{3 mm}
\\(ik_{x_{1}})c_{k_1}^{(1)}&(ik_{x_{3}})c_{k_3}^{(1)}&(ik_{x_{5}}c_{k_5}^{(1)}&(ik_{x_{2}})c_{k_2}^{(1)}&(ik_{x_{4}})c_{k_4}^{(1)}&(ik_{x_{6}})c_{k_6}^{(1)}\vspace{3 mm}
\\(ik_{x_{1}})c_{k_1}^{(2)}&(ik_{x_{3}})c_{k_3}^{(2)}&(ik_{x_{5}})c_{k_5}^{(2)}&(ik_{x_{2}})c_{k_2}^{(2)}&(ik_{x_{4}})c_{k_4}^{(2)}&(ik_{x_{6}})c_{k_6}^{(2)}\vspace{3 mm}
\\(k_{x_{1}}^{2})c_{k_1}^{(1)}&(k_{x_{3}}^{2})c_{k_3}^{(1)}&(k_{x_{5}}^{2})c_{k_5}^{(1)}&(k_{x_{2}}^{2})c_{k_2}^{(1)}&(k_{x_{4}}^{2})c_{k_4}^{(1)}&(k_{x_{6}}^{2})c_{k_6}^{(1)}\vspace{3 mm}
\\(k_{x_{1}}^{2})c_{k_1}^{(2)}&(k_{x_{3}}^{2})c_{k_3}^{(2)}&(k_{x_{5}}^{2})c_{k_5}^{(2)}&(k_{x_{2}}^{2})c_{k_2}^{(2)}&(k_{x_{4}}^{2})c_{k_4}^{(2)}&(k_{x_{6}}^{2})c_{k_6}^{(2)}\vspace{3 mm}
\end{pmatrix},
\end{equation}
\begin{equation}
Z_{n}=\text{diag}
\begin{pmatrix}
e^{ik_{x_{1}}nd}\hspace{5 mm}e^{ik_{x_{3}}nd}\hspace{5 mm}e^{ik_{x_{5}}nd}\hspace{5 mm}e^{ik_{x_{2}}nd}\hspace{5 mm}e^{ik_{x_{4}}nd}\hspace{5 mm}e^{ik_{x_{6}}nd}
\end{pmatrix}.
\end{equation}
\end{widetext}
If we introduce $\mathbf{T}$ as the total transfer matrix, then the relation between scattering coefficients of the  incoming electrons at $x\leq 0$ and those of the outgoing electrons at $x\geq (N-1)d$ can be written as   
\begin{equation}\label{a0aN}
a_{\alpha}^{(N)}=\sum_{\beta=1}^6 \mathbf{T}_{\alpha\beta} a_{\beta}^{(0)},
\end{equation} 
with
\begin{equation}\label{total}
\mathbf{T}=\left[T^{(N-1)}\cdots T^{(1)}T^{(0)}\right].
\end{equation} 
Also, the wave-function coefficients in Eq.(\ref{a0aN}) are given as
\begin{eqnarray}
&&a^{(0)}=(1, 0, 0, r_1, r_2, r_3), \nonumber\\
&&a^{(N)}=(t_1, t_2,  t_3, 0, 0, 0),
\end{eqnarray} 
where $r_{\gamma}$ and $t_{\gamma}$ $(\gamma =1,2,3)$ are the reflection and transmission amplitudes. 
If all six $k_{x\alpha}$ ($\alpha=1,\cdots,6$) values are real, then the transmission and reflection probabilities are obtained from $T=\sum_{\gamma=1}^3|t_{\gamma}|^2$ and $R=\sum_{\gamma=1}^3|r_{\gamma}|^2$, respectively. On the other hand, if only two $k_{x\alpha}$ values are real then the corresponding probabilities are calculated by $T=|t_1|^2$ and $R=|r_1|^2$. 

\subsection{A superlattice of line defects along $\Gamma$M($y$)}

We now consider structural ripples as a regular array of $N$ delta-function potentials (line defects) on the surface of Bi$_{2}$Te$_{3}$ in $y$ direction,  namely the ripples are parallel to $x$ axes ($\Gamma$M direction in {\bf k}-space, shown in Fig. 1). Similar to Eq. (\ref{Vx}), the potential function of such defects is given as
\begin{equation}
V(y)=\sum_{n=0}^{N-1}U\delta(y-nd).
\end{equation}
In this case, for given values of $E_{\bf k}$ and $k_x$, the characteristic equation in terms of $k_y$ is of order four and in general has four roots. As can be seen in Eq. (\ref{Hamiltonian}), the Hamiltonian is not symmetric under $p_x\longleftrightarrow p_y$ transformation. Therefore, the new boundary conditions corresponding to the eigenvalue equation
$\{H(k_{x},-i\partial_{y})+\sum_{n=0}^{N-1}U\delta(y-nd)\}\Psi_{\bf k}^{(n)}({\bf r})=E\Psi_{\bf k}^{(n)}({\bf r})$,
that is a second-order partial differential equation with respect to $y$, are given as
\begin{eqnarray}\label{boundry2}
\Psi_{\bf k}^{(n)}({\bf r}){|_{(x,y=nd)}}&=&\Psi_{\bf k}^{(n+1)}({\bf r}){|_{(x,y=nd)}}, \\
\partial_{y}(\Psi_{\bf k}^{(n+1)}({\bf r})&-&\Psi_{\bf k}^{(n)}({\bf r})){|_{(x,y=nd)}}\nonumber\\
&&=-\kappa\eta\sigma_{z}\Psi_{\bf k}^{(n+1)}({\bf r}){|_{(x,y=nd)}}, \nonumber 
\end{eqnarray}
where $\kappa=v/3k_{x}\lambda$. The wave function in the region $(n-1)d<y<nd$ is written as 
\begin{equation}\label{Psiy}
\Psi_{\bf k}^{(n)}({\bf r})=\sum_{\beta=1}^4 b_{\beta}^{(n)}\widetilde{\psi}_{k_{\beta}}({\bf r}),
\end{equation}
with
\begin{eqnarray}
b^{(0)}=(1, 0, r_1, r_2),\hspace{5 mm} 
b^{(N)}=(t_1, t_2, 0, 0).
\end{eqnarray}

Inserting Eq. ({\ref{Psiy}}) in the boundary conditions of Eq. ({\ref{boundry2}}), we obtain a local  $4\times 4$ transfer matrix, $T^{(n)}=(Z_n^{-1}A^{-1}BZ_n)$ where $A$, $B$ and $Z_{n}=\text{diag}
\begin{pmatrix}
e^{ik_{y_{1}}nd}\hspace{5 mm}e^{ik_{y_{3}}nd}\hspace{5 mm}e^{ik_{y_{2}}nd}\hspace{5 mm}e^{ik_{y_{4}}nd}
\end{pmatrix}$ are $4\times 4$ matrices. Therefore, when all modes are propagating, the transmission and reflection probabilities can be obtained from $T=\sum_{\gamma=1}^2|t_{\gamma}|^2$ and $R=\sum_{\gamma=1}^2|r_{\gamma}|^2$, respectively. However, the probabilities are calculated from $T=|t_1|^2$ and $R=|r_1|^2$, if there exist only two propagating modes. 

\begin{figure}
\centerline{\includegraphics[width=0.47\textwidth]{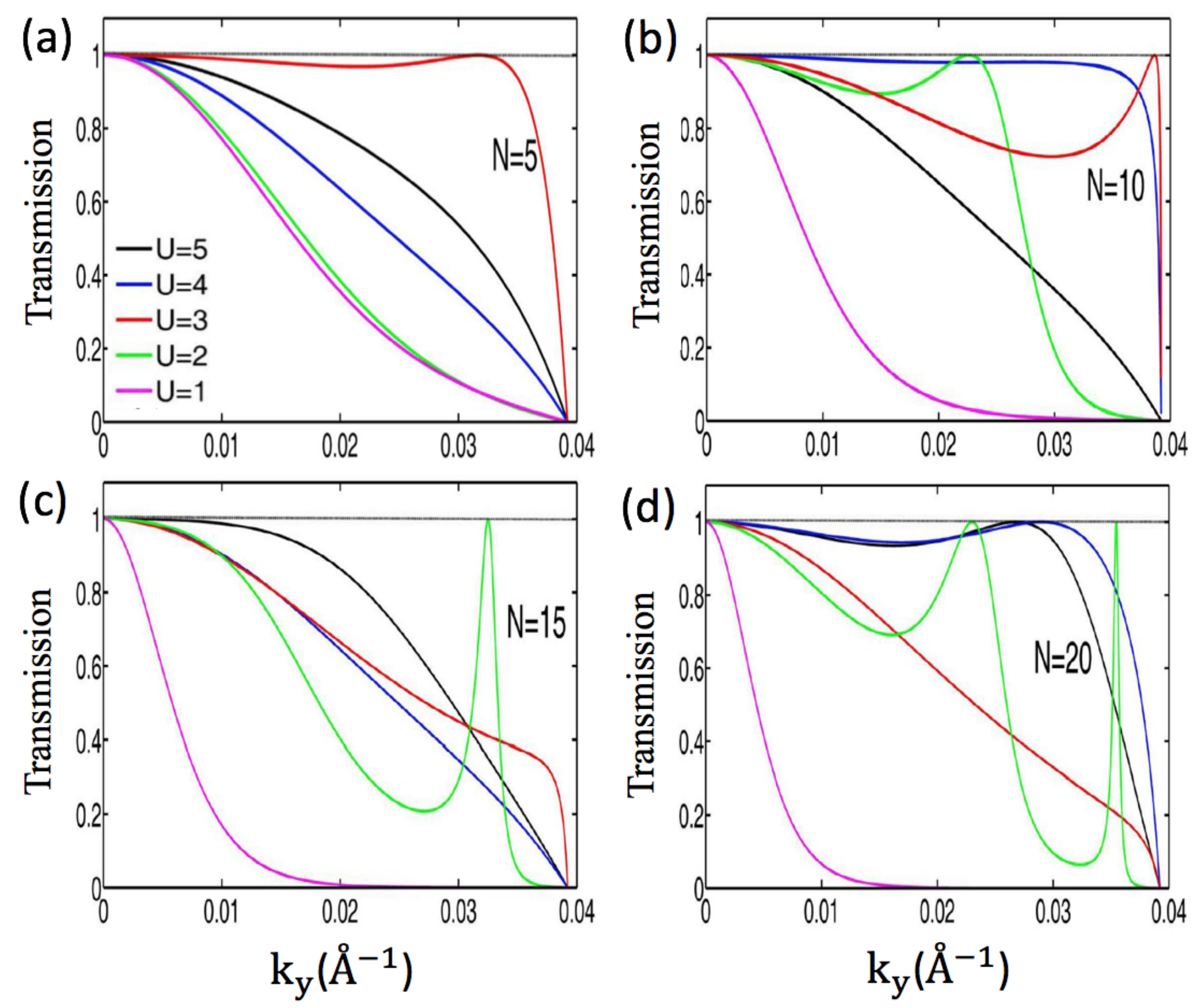}}
 \caption{Calculated transmission probabilities versus $k_y$ at energy $E=100$ meV for different number of defects, $N$, running along $y$ ($\Gamma$M) direction and various potential strengths $U$, measured in units of eV$\cdot${\AA}. The lattice constant $d$ is 10 \AA.}
\label{fig:3}
\end{figure}

\begin{figure}
\centerline{\includegraphics[width=0.47\textwidth]{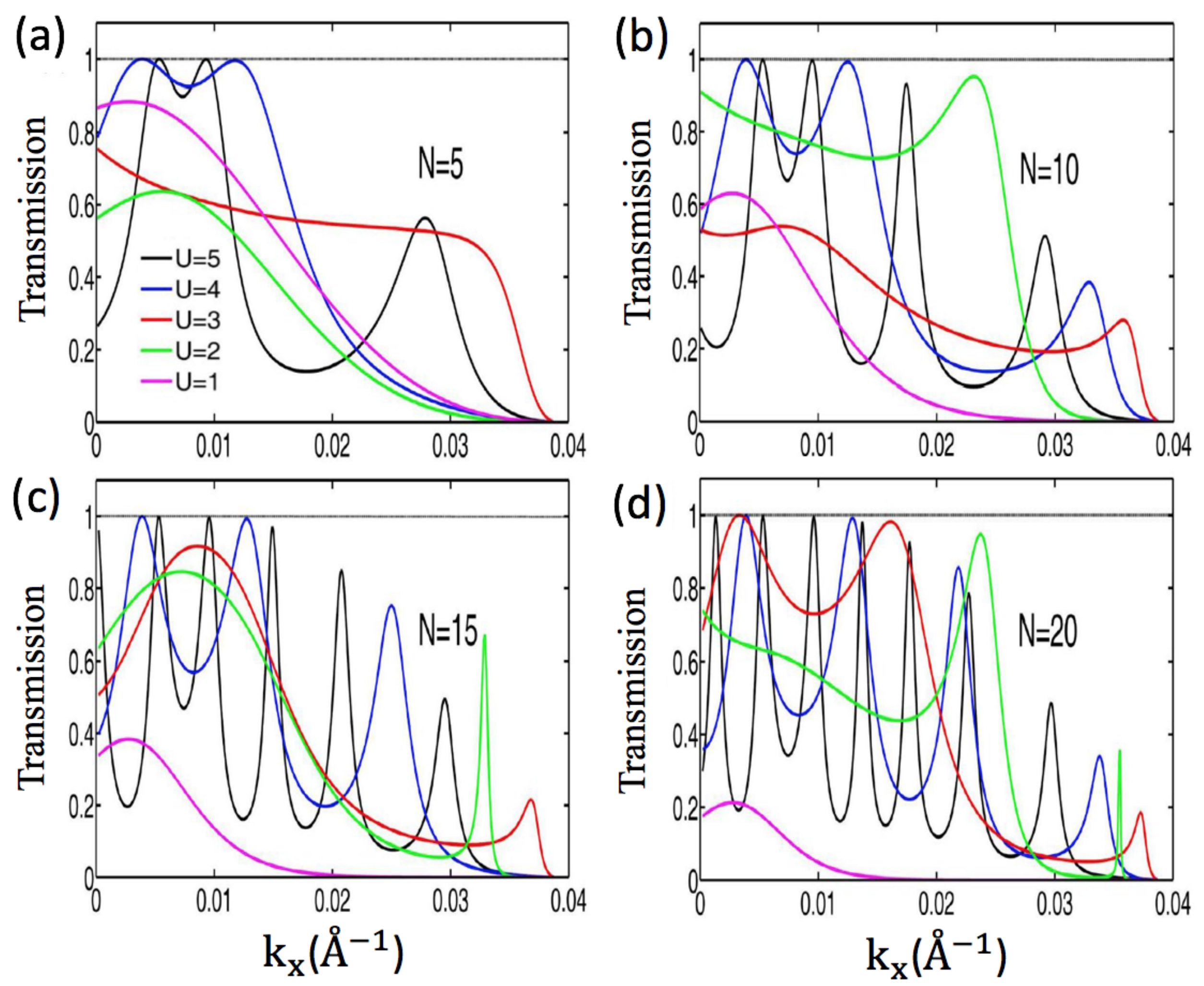}}
\caption{Calculated transmission probabilities versus $k_x$ at energy $E=100$ meV for different number of defects, $N$, running along $x$ ($\Gamma$K) direction and various potential strengths $U$, measured in units of eV$\cdot${\AA}. The lattice constant $d$ is 10 \AA.}
\label{fig:4}
\end{figure}

Finally, using the transmission probability $T(E,\theta)$, the electrical conductance of  the topological insulators in the $\Gamma$K($x$) direction can be calculated from \cite{li2013-JAP}
\begin{equation}
G/G_0=\int_{0}^{\pi/2}{T(E,\theta)}\cos\theta d\theta,
\end{equation}
where $\theta=\arctan(k_y/k_x)$ is the electron incidence angle and $G_0$ is the unit of conductance.
   
\section{RESULTS AND DISCUSSION}
To explore the dependence of incident energy and wave vector of electrons on transport through the superlattice of localized defects, we have shown in Fig. 3 the transmission probabilities in terms of $k_{y}$ at incident energy $E=100$ meV ($<E_{c}$) with different numbers of defects, extended along the $y$ direction. The perfect transmission at $k_{y}=0$ indicates the absence of backscattering for incident electrons. This effect which is known in relativistic field theory as Klein tunneling \cite{Klein-1929,Katsnelson2006}, happens for all $N$ and $U$ values. When $k_{y}$ (angle of incidence) is increased for any value of $U$, the transmission probability changes continuously. Also, for any value of $k_{y}$ (i.e., fixed angle of incidence), $T$ changes alternatively in terms of $U$. These are single barrier features found in single barrier structures on graphene \cite{Ramezani,Agrawal-EPJB} and are seen here as well.  

The resonant peaks seen in Fig. 3 can be attributed to constructive interferences due to multiple reflection of electron waves between the line defects. The position and the number of resonant peaks, as already seen in double-barrier structures on graphene \cite{Ramezani,Agrawal-EPJB}, depend not only on the strength of $U$, but also on the acquired phase ($k_{x}d$) of propagating waves between two line defects, i.e., $d$ and the wave vector component $k_{x}$, which is $E$ and $\theta$ dependent. It is clear that at incident energy $E=100$ meV, only electrons with $k_{y}< 0.04${\ \AA}$^{-1}$ can propagate through the superlattice (see Fig. 3). The zero transmission probability at $k_{y}=0.04${\ \AA}$^{-1}$ is trivial because the incident electrons at such wave vector can travel only parallel to the line defects without crossing the barriers.          

Now we consider the situation in which the defects are extended along the $x$ direction. For this case, the transmission probabilities as a function of $k_{x}$ at energy $E=100$ meV are depicted in Fig. 4. Unlike the perfect transmission at $k_{y}=0$ in Fig. 3, here the transmission probability is not perfect at $k_{x}=0$ (normal incidence). This is because for nearly normal incidence, the existence of localized decaying modes at the defects can act as magnetic delta-function barrier for propagating electrons, leading to finite reflections \cite{an2012-PRB}. By changing the direction of propagation, the wave vector changes from $k_{y}$ to $k_{x}$. This leads to a different acquired phase of propagating waves between the defects and hence, the number and the position of resonances show a drastic change, compared to those in Fig. 3. The change in the resonances by varying the propagation direction from $k_{y}$ to $k_{x}$ is coming from the fact that Eq. (2) in term of $k_y$ is of order four, whereas it is of order six in term of $k_x$, causing a remarkable change in the transmission spectrum. The zero transmission which is also trivial at $k_{x}=0.04${\ \AA}$^{-1}$ is due to the incident electrons parallel to the barriers.

By comparing Figs. 3 and 4, we find that at low energies, the electron transmission through the localized defects on the surface of topological insulator Bi$_2$Te$_3$ depends strongly on the direction in which the line defects extend. Moreover, the none-zero transmission probabilities for electrons with $k_{x(y)}<0.04${\ \AA}$^{-1}$ in both figures confirms a circular CEC at low energies, even in the presence of warping term in the total Hamiltonian. The difference in the transport properties of electrons through the line defects in $x$ and $y$ directions comes from different boundary conditions Eqs. (\ref{boundry1}) and (\ref{boundry2}), and also Eq. (\ref{EK}) which gives the roots of $k_{x}$ and $k_{y}$. All these differences are originated from the asymmetry of Hamiltonian under $k_{x}\longleftrightarrow k_{y}$ transformation. The discrepancy in the transmission spectra in both directions becomes significant as the potential strength is increased.
\begin{figure}
\centerline{\includegraphics[width=0.47\textwidth]{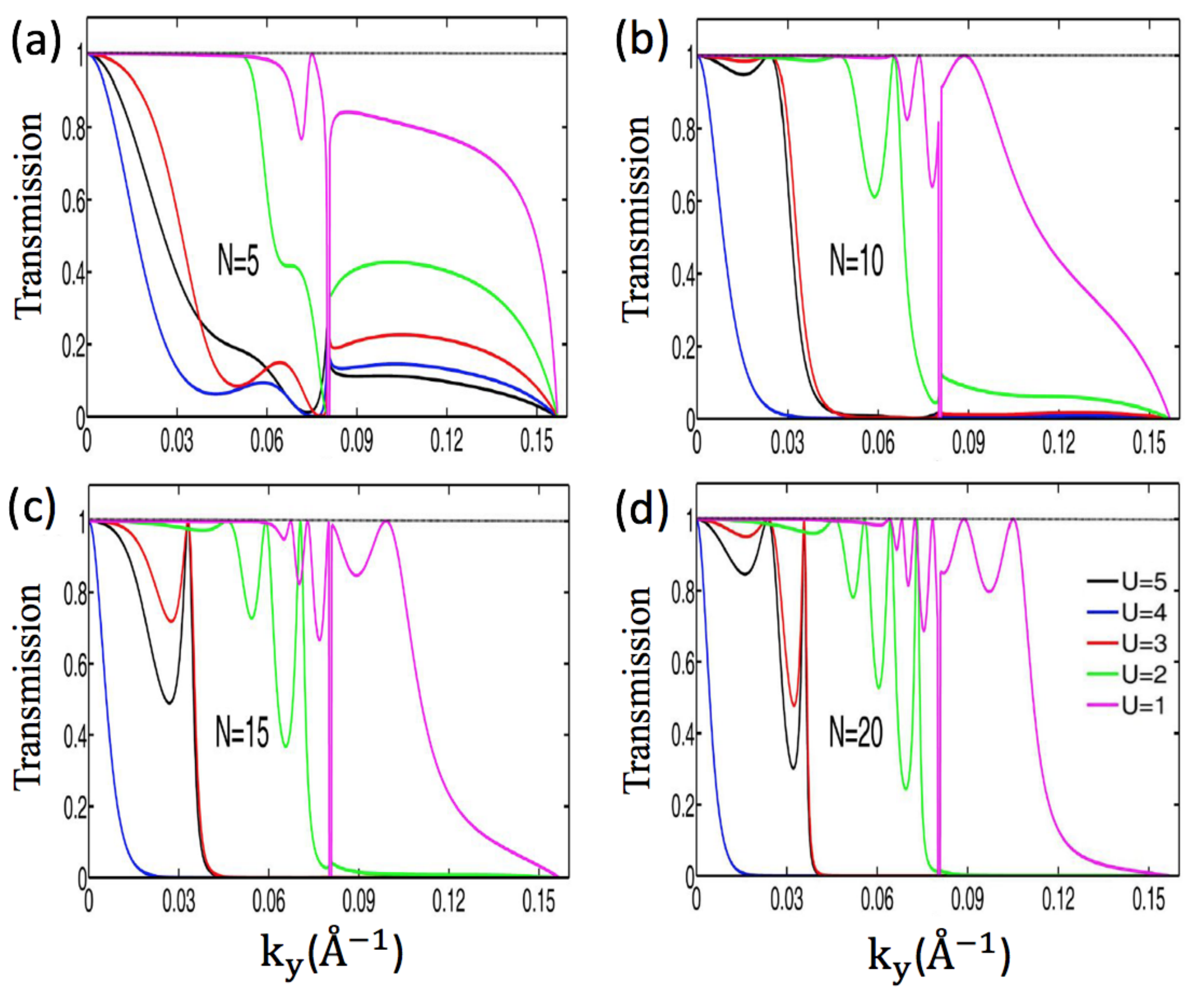}}
\caption{Calculated transmission probabilities versus $k_y$ at incident energy $E=400$ meV for different number of defects, $N$, running along $y$ ($\Gamma$M) direction and various potential strengths $U$, measured in units of eV$\cdot${\AA}. The lattice constant $d$ is 10 \AA.}
\label{fig:5}
\end{figure}
\begin{figure}
\centerline{\includegraphics[width=0.47\textwidth]{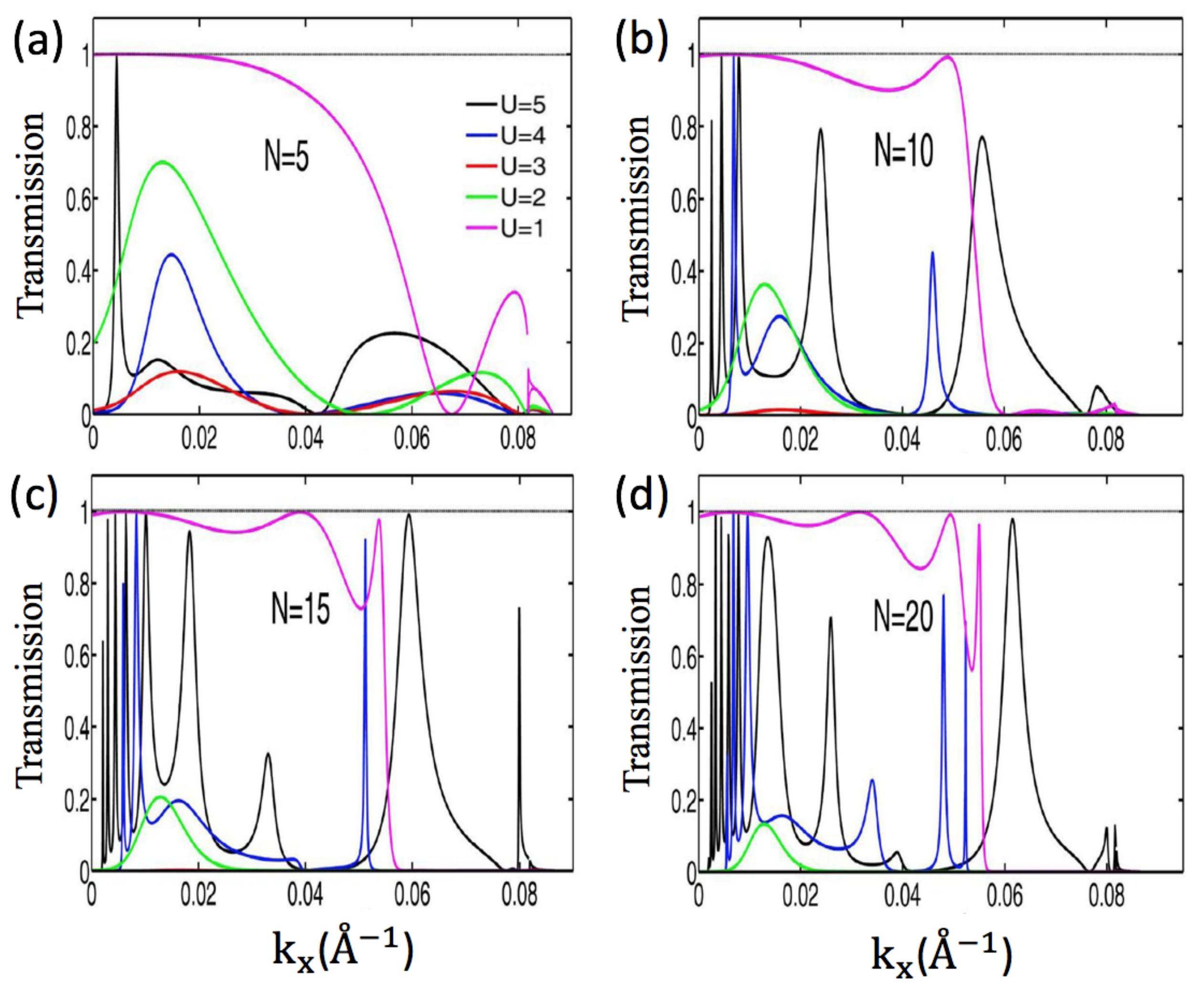}}
\caption{Calculated transmission probabilities versus $k_x$ at incident energy $E=250$ meV for different number of defects, $N$, running along $x$ ($\Gamma$K) direction and various potential strengths $U$, measured in units of eV$\cdot${\AA}. The lattice constant $d$ is 10 \AA.}
\label{fig:6}
\end{figure}

The incident electrons with $E>E_{c}$, however, are greatly modulated by the warped energy contour. We have depicted in Figs. 5 and 6 the transmission probabilities of electrons at $E=400$ meV and $E=250$ meV through a superlattice of line defects, extended along $y$ and $x$ directions, respectively. In this case, the incident energy is greater than the critical value in both directions and the CEC becomes clearly warped, as shown in Fig. 1. In fact, at high energies the change in $k_{x,y}$ interval increases, leading to an increase in resonant peaks. In addition, in some $k_{x,y}$ intervals, the number of reflected and transmitted waves for each single barrier enhances, due to the presence of warping effect. These can cause an increase in the number of maxima in the transmission spectra.  In Fig. 5, the transmission probabilities exhibit three branches due to six real roots in the momentum interval 0.08 {\AA}$^{-1}\leq k_{y}\leq$0.081 {\AA}$^{-1}$, as a result of strong warping effect. Therefore, extreme points with $T=1$ or 0 appear (see the sharp drop in $T$ at $k_{y}=0.08$ {\AA}$^{-1}$). This means that in this momentum interval, incident electrons can be perfectly transmitted or totally reflected. It is worth mentioning that the increase in the number of line defects does not affect the momentum interval in which $T$ drops and rises sharply. Also, the increase in the incident energy value enhances the resonant peaks as the number of defects is increased. 
Similar to the situation shown in Fig. 3, the Klein tunneling happens at $k_{y}=0$. Moreover, the zero transmission (total reflection) at $k_{y}=0.157$ {\AA}$^{-1}$ corresponds to an incident wave propagating along the line defects without being scattered.

On the other hand, as can be seen in Fig. 6, the transmission spectrum in the $x$ direction is different with that in the $y$ direction. In this case the transmission probability shows two branches due to four real roots in the momentum interval 0.082 {\AA}$^{-1}\leq k_{x}\leq$0.086 {\AA}$^{-1}$ in which the incident wave tends to become tangent to the defect. The total reflection at $k_{x}=0.082$ {\AA}$^{-1}$ corresponds to an incident wave propagating parallel to the defects, while the total reflection occurs at $k_{x}=0.086$ {\AA}$^{-1}$ corresponding to a point on the sharp corner of CEC (see Fig. 1) in which the group velocity of electrons in $y$ direction becomes zero \cite{an2012-PRB}. By increasing the number of line defects and/or the potential strength, the transport through the superlattice varies drastically. For instance, as $N$ increases, the transmission spectrum through the line defects with the potential strength $U=3$ eV$\cdot${\AA} vanishes (compare the red curves in Fig. 6(a)-(d)) due to a strong destructive interference scattering from the line defects, while the finite propagation occurs for the higher $U$ values and the corresponding number of resonant peaks enhances. In addition, similar to the situation shown in Fig. 4, the transmission is not perfect at $k_{x}=0$ (normal incidence). We note that in the case of barriers along $x$ axis ($y$ axis), the incident direction is closer to the tangent (normal) line of the barriers for the momentum values at which multiple scattering occurs, regardless of the electron energy (see Fig. 1). This reveals that the electron transport in the presence of structural ripples on the surface of topological insulators is direction dependent as a result of warping effect. 

Based on the obtained results from Figs. (3) to (6), one can conclude that at low energies where the CEC has a circular shape due to the weak warping effect (see Fig. 1), the dominant contribution of the resonant peaks in the transmission spectrum comes from the quantum interference between the incident and reflected propagating electrons. At higher energies $E >E_{c}$, however, the interference between the incident and reflected propagating waves and also the multiple scattering processes on the CEC due to the strong warping effect, contribute in the transmission spectrum.
    
\begin{figure}
\centerline{\includegraphics[width=0.47\textwidth]{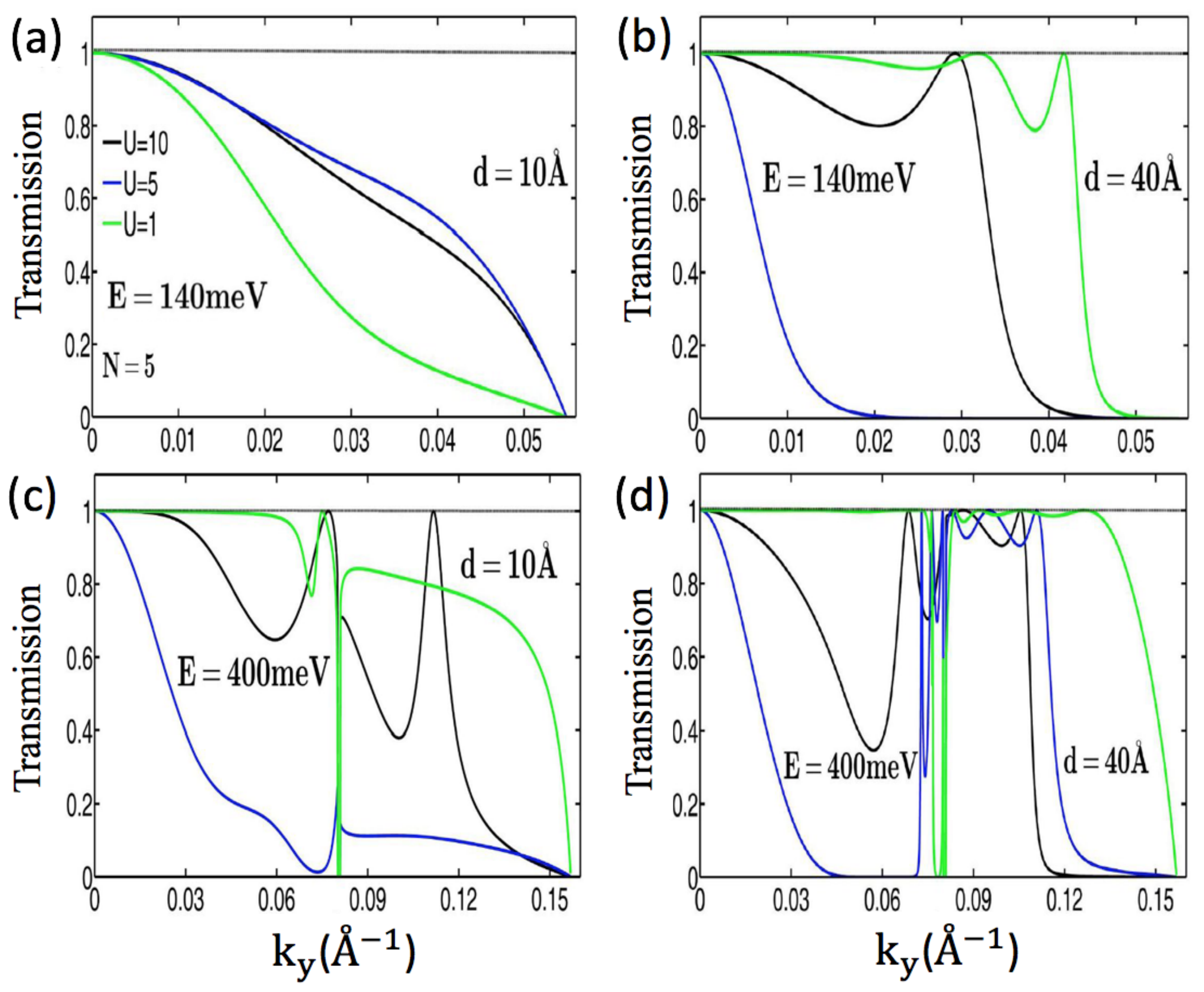}}
\caption{Calculated transmission probabilities versus $k_y$ for two different $d$ and $E$ values. The number of potential barriers is fixed at $N=5$. $U$ is measured in units of eV$\cdot${\AA}.}
\label{fig:7}
\end{figure}

The spatial separation between consecutive potential ripples can be a key factor in direct correlation between the local potentials. Such a separation is called the lattice constant $d$ in our theory. Therefore, to investigate the importance of distance between the potential barriers, we have depicted in Fig. 7  the transmission probabilities versus momentum $k_{y}$ for fixed number of localized defects running along $y$ direction but different $d$ values.  As shown in Fig. 7(a) for electrons with incident energy $E=140$ meV ($< E_{cx}$) and lattice constant $d=10$ {\AA}, by increasing the momentum $k_{y}$, the transmission probabilities gradually suppress, whereas some resonant peaks are induced in the $T$ spectrum at $d=40$ {\AA} (see Fig. 7(b)). The appearance of resonances depends on the potential strength $U$, affecting the location of resonant energies between the potential barriers. Therefore, the increase in the number of resonant peaks does not represent a linear relation with the increase in the $U$ values. In contrast, the $T$ spectrum for electrons with incident energy $E=400$ meV ($>E_{cx}$) and lattice constant $d=10$ {\AA} demonstrates some resonant peaks for all $U$ values, as shown in Fig. 7(c). Such resonances are intensified in Fig. 7(d) for the structure with $d=40$ {\AA}, due to the expansion of acquired phase between potential barriers on the surface of topological insulator. 

\begin{figure}
\centerline{\includegraphics[width=0.5\textwidth]{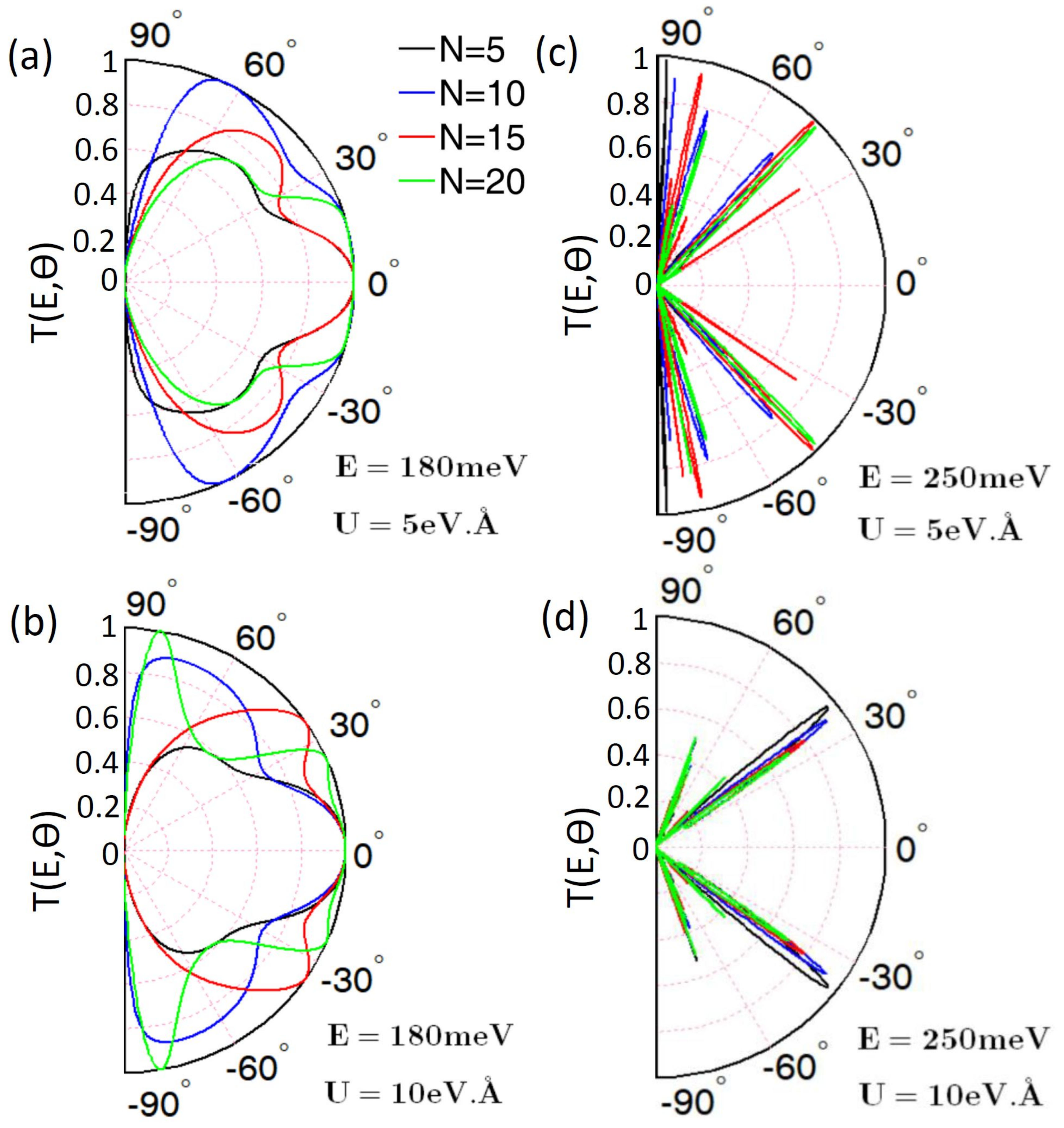}}
\caption{Calculated transmission probabilities in terms of electron incidence angle $\theta$ at energies (a)-(b) $E=180$ meV and (c)-(d) $E=250$ meV for different number of line defects. The diagrams (a) and (b) correspond to the superlattices with lattice constant $d=9$ {\AA} running along the $y$ direction, while (c) and (d) correspond to the superlattices with lattice constant $d=10$ {\AA} running along the $x$ direction.}
\label{fig:figure8}
\end{figure}

According to the above discussion, the appearance of resonant peaks in the transmission spectra is dependent on the scattering strength $U$ and the number of defects $N$, the distance $d$ between two adjacent defects, electron energy $E$, and the angle of incidence $\theta$. By varying these parameters, the number of resonances changes irregularly and one cannot always expect a systematic trend in the transmission spectrum. Nevertheless, by tuning the parameters and changing one of them, regular features can be observed. For instance, Fig. 6 for $U=1$ eV$\cdot$\AA\ shows that by increasing the number $N$ of line defects, the number of maxima (Fabry-Perot resonances) increases, due to the multiple reflections, which can be understood from double barrier structures on graphene \cite{Agrawal-EPJB}. 

To examine the transmission probabilities of electrons at different incident angles, $\theta$, we have shown $T$ versus $\theta$ in Figs. 8(a)-(b) and 8(c)-(d) for a series of line defects extending along $y$ and $x$ directions, respectively. Interestingly, we can see in Fig. 8(a)-(b) perfect transmissions for normally incident electrons ($\theta=0$), regardless of the potential strengths and the number of ripples in $y$ direction, which is a typical signature of Klein tunneling \cite{Katsnelson2006,Hassan2017}. Moreover, the transmission profile is symmetrical about ($\theta=0$) due to the absence of an external magnetic field along $x$ direction \cite{Arabikhah2019}. Note that although a magnetic field can shift the perfect transmission from normal incidence to an off-normal angle ($\theta\neq 0$) when the magnetic field is along the $x$ direction, time-reversal symmetry is not broken and the system remains conductive for all surface electron energies \cite{Arabikhah2019}. The change in the potential strength can shift the incident angles with perfect transmission. Also, the change in the $U$ values may extend the non-zero transmission over a wider range of incident angles, as shown in Fig. 8(b). This reveals that the propagation of electrons with incident energy $E$ in a superlattice of $N$ line defects running along $y$ direction can be controlled by the incident angle of electrons and the potential strength. On the other hand, in Figs. 8(c)-(d), the transmission profiles of electrons at incident energy $E=250$ meV in a superlattice of line defects along $x$ direction exhibit only strong spikes as a result of resonant peaks, seen in the electron transmission spectra in this direction, regardless of the number of ripples. The transmission of electrons is forbidden in a wide range of $\theta$, especially for tangential incident electrons, indicating that $T(E,\theta)$ for electrons propagating through line defects extending along $x$ direction is quite different than that seen in Figs. 8(a)-(b). In Figs. 8(c) and (d), the warping effect is significant. Moreover, the angle of incidence $\theta$ is measured with respect to the normal line. Therefore, for the interval $0^{\circ}\le\theta\le 30^{\circ}$ in which the incident wave tends to become tangent to the barrier, $T$ approaches zero. Also, at $\theta=60^{\circ}$ (the sharp corner of CEC) and $\theta=30^{\circ}$ (the extreme point between two sharp corners of CEC) the transmission probability is zero, because the group velocity becomes zero, i.e, $v_{y}=(\frac{\partial E}{\partial k_{y}})_{k_{x}}=-v_{x}(\frac{\partial k_{x}}{\partial k_{y}})_{E}=0$ \cite{an2012-PRB}.

\begin{figure}
\centerline{\includegraphics[width=0.5\textwidth]{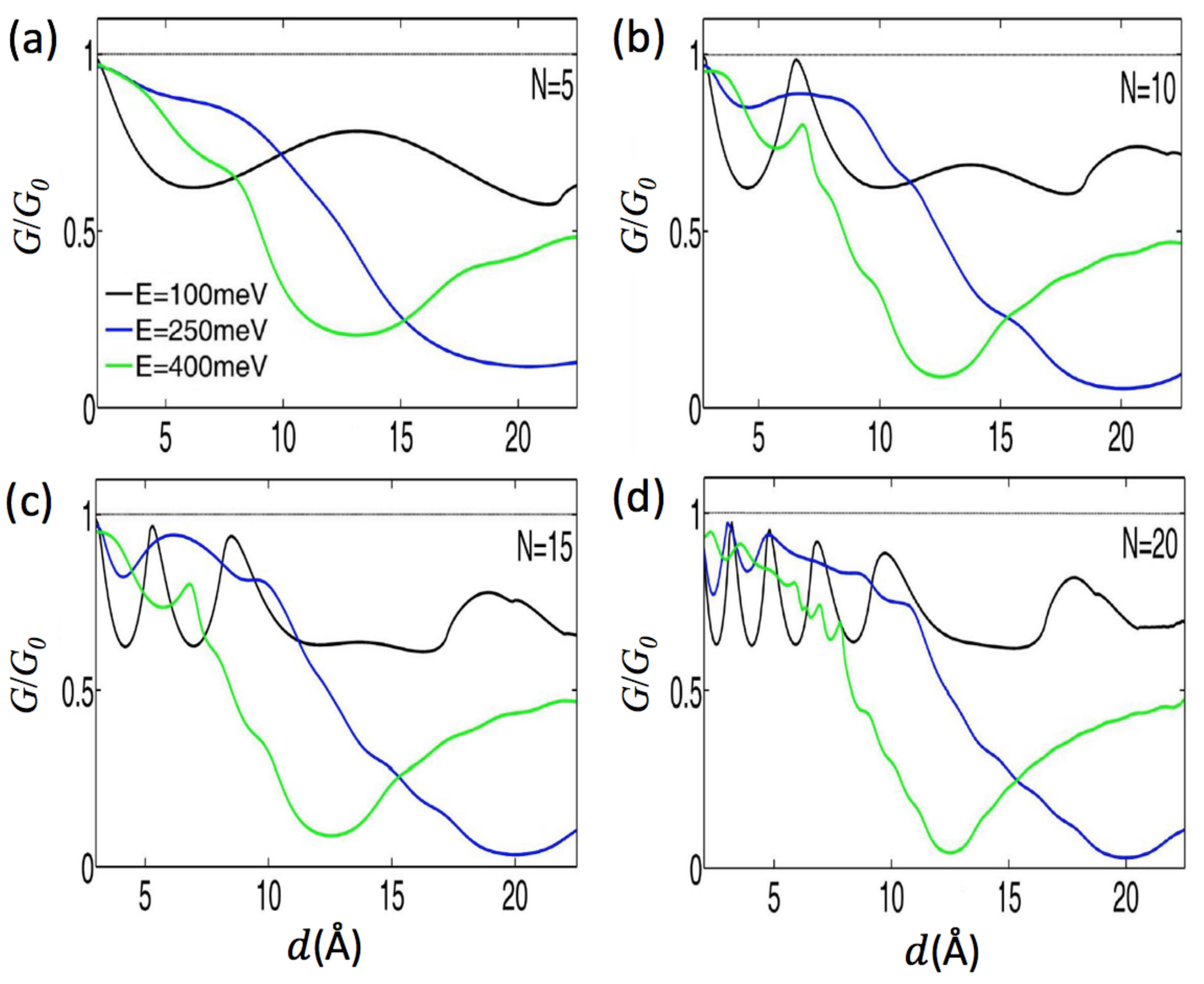}}
\caption{Calculated conductance in superlattices of line defects with potential strength $U=5$ eV$\cdot${\AA}, consisting of (a) $N=5$, (b) $N=10$, (c) $N=15$, and (d) $N=20$ defects in terms of lattice constant $d$.}
\label{fig:9}
\end{figure}

As mentioned above, such a feature comes from the asymmetry of Hamiltonian under $k_{x}\longleftrightarrow k_{y}$ transformation and originates from the hexagonal warping effect shown in Fig. 1. This suggests that the influence of periodic ripples on transport properties of electrons on the surface of topological insulators is strongly dependent on the direction in which the ripples are created. By forming such controlled ripples, one can tune the electronic properties of Dirac fermions in the surface of topological insulators \cite{okada2012-NAT}.
    
Finally, we have shown in Fig. 9 the electrical conductance in superlattices of line defects versus $d$ for various number of defects. By increasing the number of line defects, the conductance exhibits oscillatory behavior at low $d$ values. These oscillations are significant at incident energy $E=100$ meV, while the amplitude of the oscillations decreases as $E$ is increased. The conductance remains above 0.5$G_{0}$ at $E=100$ meV in the given $d$ interval, because the quantum interference between the incident and reflected propagating electrons is the only factor in the appearance of transmission resonances at low energies \cite{seo2010-NAT}. In contrast, the conductance may approach 0.05$G_{0}$ at high energies due to the multiple scattering processes on the CEC as a result of strong warping effect. These results clearly reveal that the spatial separation between line defects, which can be experimentally created by strain via a piezoelectric crystal on the surface of topological insulators \cite{okada2012-NAT}, is crucial in controlling the electric conductance. 
                                           
\section{CONCLUSIONS}

In summary, using a transfer matrix method, we have examined theoretically the influence of periodic line defects, represented by delta-function potential barriers, and hexagonal warping effect on electronic transport on the surface of topological insulators. We have shown that the electron transmission and the resonant states are strongly dependent on the number and strength of potential barriers and the direction in which the defects extend. The warped-energy contour may cause multiple scattering processes at high energies, whereas the electron transmission is mostly affected by the quantum interference between the incident and reflected electrons at low energies. Moreover, the electrical conductance versus the spatial separation between the potential barriers exhibits some oscillations as the number of defects is increased. This suggests that the surface electronic properties of topological insulators can be controlled by creation of structural ripples in an appropriate direction and also adjusting the energy and incident angle of electrons. The preset formalism can also be utilized to study ordered and/or disordered distributions of an alternating set of line defects on the surface of topological insulators.

\clearpage
\onecolumngrid
\setcounter{equation}{0}
\setcounter{figure}{0}

\setcounter{page}{1}
\makeatletter
\renewcommand{\theequation}{S\arabic{equation}}
\renewcommand{\thefigure}{S\arabic{figure}}

\end{document}